\documentclass[letterpaper, 10 pt, conference]{ieeeconf}  

\IEEEoverridecommandlockouts                              

\usepackage{amsmath} 
\usepackage{amssymb}  
\usepackage{graphicx}
\usepackage{tikz}
\usepackage{hyperref} 
\usepackage{pdfpages} 
\usepackage{xcolor}
\usepackage{bm}

\title{\LARGE \bf
A Tricycle Model to Accurately Control an\\ Autonomous Racecar with Locked Differential
}

\author{Ayoub Raji$^{1,2,*}$, Nicola Musiu$^{1,*}$,  Alessandro Toschi$^{1}$, Francesco Prignoli$^{1}$, Eugenio Mascaro$^{1}$,\\ Pietro Musso$^{1}$, Francesco Amerotti$^{3}$, Alexander Liniger$^{4}$, Silvio Sorrentino$^{1}$, Marko Bertogna$^{1}$
\thanks{$^{1}$University of Modena and Reggio Emilia, Italy \newline
        {\tt\small \{firstname.lastname\}@unimore.it},\newline
        {\tt\small \{291709;304753\}@studenti.unimore.it},
        }%
\thanks{$^{2}$University of Parma, Italy}
\thanks{$^{3}$Hipert srl, Italy
\newline
        {\tt\small francesco.amerotti@hipert.it}}%
\thanks{$^{4}$Computer Vision Lab, ETH Zurich, Switzerland \newline
        {\tt\small alex.liniger@vision.ee.ethz.ch}}%
\thanks{$*$The authors contributed equally.}
}

\newcommand\copyrighttext{%
\footnotesize \copyright 2023 IEEE. Personal use of this material is permitted. Permission from IEEE must be obtained for all other uses, in any current or future media, including reprinting/republishing this material for advertising or promotional purposes, creating new collective works, for resale or
redistribution to servers or lists, or reuse of any copyrighted component of this work in other works.}
\newcommand\copyrightnotice{%
\begin{tikzpicture}[remember picture,overlay]
\node[anchor=north,yshift=-20pt] at (current page.north) {\fbox{\parbox{\dimexpr\textwidth-\fboxsep-\fboxrule\relax}{\copyrighttext}}};
\end{tikzpicture}%
}

\begin{document}

\maketitle
\thispagestyle{empty}
\pagestyle{empty}

\begin{abstract}

In this paper, we present a novel formulation to model the effects of a locked differential on the lateral dynamics of an autonomous open-wheel racecar. The model is used in a Model Predictive Controller in which we included a micro-steps discretization approach to accurately linearize the dynamics and produce a prediction suitable for real-time implementation.
The stability analysis of the model is presented, as well as a brief description of the overall planning and control scheme which includes an offline trajectory generation pipeline, an online local speed profile planner, and a low-level longitudinal controller.
An improvement of the lateral path tracking is demonstrated in preliminary experimental results that have been produced on a Dallara AV-21 during the first Indy Autonomous Challenge event on the Monza F1 racetrack. Final adjustments and tuning have been performed in a high-fidelity simulator demonstrating the effectiveness of the solution when performing close to the tire limits.
\end{abstract}
\copyrightnotice
\begin{keywords}
Autonomous racing, model predictive control, tricycle model, single-track model, vehicle modeling
\end{keywords}
\section{INTRODUCTION}
\label{introduction}

The vehicle models used within model-based motion planning and control algorithms for autonomous driving depend on a trade-off between simplicity and accuracy, which is closely tied to the specific application. The most common models in the literature are the point-mass model, kinematic single-track model, dynamic single-track model, double-track model, and multi-body model \cite{commonroad}. In urban scenarios, where motion can be identified through purely geometric approaches, due to the minimal slip angles, it is generally sufficient to use a kinematic single-track model \cite{kong}\cite{paden}.

In more complex maneuvers that produce not negligible lateral accelerations, the dynamic single-track model is commonly used. Dynamic effects, such as tire forces, and additional states such as lateral velocity, are considered. In \cite{kirstin}\cite{kapania}, the model is used in a feedback-feedforward steering control scheme guaranteeing accuracy and stability. In more recent works, the Model Predictive Control (MPC) using the single-track model has become widely used \cite{stano}. Several researchers demonstrated the effectiveness of this approach to control sport road cars close to the limit of handling exploiting more accurate tire models to represent the lateral forces \cite{zhang}\cite{dallas}\cite{talbot}. We implemented a similar solution competing as the TII Unimore Racing team to the Indy Autonomous Challenge (IAC{\footnote{\href{https://www.indyautonomouschallenge.com/}{https://www.indyautonomouschallenge.com/}}), an autonomous racing competition among universities from all around the world. Relying on the modeling of a single-track model exploited in an MPC, we achieved a top speed of 270 km/h and a maximum lateral acceleration of up to 25 m/s\textsuperscript{2} in the first events on oval tracks \cite{raji}.
To get a better overview of the methods used in the autonomous racing literature we refer to \cite{betz}. 
\begin{figure}[t]
	\centering
	\includegraphics[clip, trim=5cm 10cm 5cm 0cm, width=1\columnwidth]{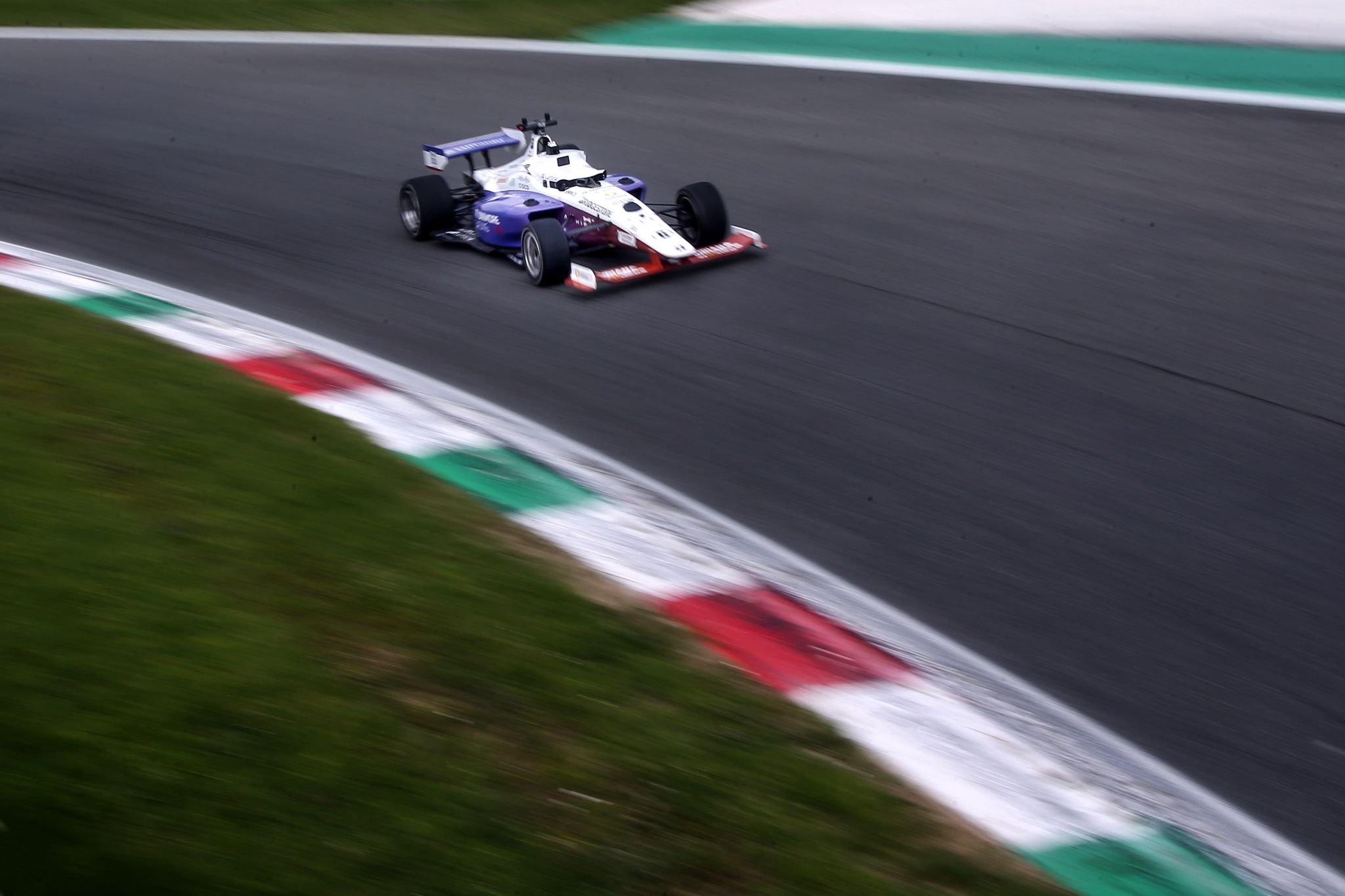}
    \caption{Dallara AV-21 - TII UNIMORE Racing during the Indy Autonomous Challenge at MIMO 2023 in the Monza F1 circuit. \copyright Autodromo Nazionale Monza}
	\label{fig:car}
\end{figure}

Despite the overall effectiveness of the bicycle model, its classical formulation has some limitations for an open-wheel racecar, e.g. assuming an open-differential. This leads to neglecting the contribution of longitudinal forces on the yaw moment.
Indeed, on a road course like the Monza F1 racetrack, where the latest event of the IAC has been held, this model presents limitations because it is not able to distinguish between tight curves and wide-radius curves. Consequently, it cannot reproduce any dynamic behavior, such as nose-in/nose-out \cite{bergman}, caused by a locked differential. Potentially, the parameters of the constitutive equations can be tuned to enhance optimal accuracy in only one of the two scenarios mentioned above. However, this can lead to highly unbalanced axle characteristics (high rear cornering stiffness), resulting in numerical instability of the model at low speed.

In response to these challenges, the main contributions of this paper are as follows:
\begin{itemize}
\item A three-wheel model (tricycle model) has been developed to account for the dependence on the path curvature. By estimating the longitudinal forces at the rear axle and, consequently, the contribution on the yaw moment, we can anticipate a possible nose-out resulting in an improvement of the vehicle modeling and therefore the trajectory tracking. To the best of our knowledge, the model proposed is new in the literature. 

\item Due to the characteristics of the system at low speed, a kinematic model has been used for low-speed maneuvers. To keep this speed threshold low, a stability analysis of the dynamic model has been performed finding the suitable discretization step time needed to keep the model stable, and implementing a micro-step discretization approach in the MPC to make it suitable for real-time execution.
\end{itemize}

In Section \ref{vehicle_modeling} we present the tricycle model formulation and its stability analysis. The MPC formulation, including the micro-step model discretization, is described in Section \ref{mpc_sec}. Even though they do not present any novel contribution, the high-level planning scheme and the low-level longitudinal controller used in the whole planning and control scheme are briefly reported in Section \ref{planning_sec} and Section \ref{lowlevel_sec}, respectively. The model validation, preliminary experimental results, and final results of the proposed solution are presented in Section \ref{results_sec}. Conclusions and future works are discussed in Section \ref{conclusions_sec}.


\section{VEHICLE MODELING}
\label{vehicle_modeling}

The Dallara AV-21, shown in Figure \ref{fig:car}, is based on the chassis and suspension of the Indy Lights IL-15. The internal combustion engine is equipped with a turbocharger that increases the engine torque to over 500 Nm. The torque is transmitted to the rear wheels through a rigid axle (no differential is installed). For vehicle handling, this represents a significant limitation, especially on road course tracks where both tight and wide-radius curves can be found. This, combined with a stable setup choice, leads the car to exhibit a nose-out behavior on turn entry, which can become a nose-in behavior on turn exit \cite{bergman}.

\subsection{Accurate Multi-Body Model}
\label{dymola}

In a context where on-track test sessions are often limited and expensive, multi-body models can offer detailed and in-depth simulations, significantly expediting the development process and allowing engineers to virtually test various scenarios and control strategies before transitioning to expensive and risky physical tests.

An accurate vehicle model is of paramount importance in this application. It provides an efficient solution to optimize vehicle performance and to test planning and control algorithms using a reliable ground truth.
Therefore, a multi-body model of the Dallara AV-21 has been developed on the modeling tool Dymola \cite{dymola} using the VeSyMA - Motorsports libraries \cite{dempsey}. The model includes all the longitudinal, lateral, and vertical dynamics of the real system. Compared to the previous version mentioned in \cite{er-autopilot}, particular attention has been focused on modeling the suspension dampers, the tire's thermal behavior, and the powertrain inertial characteristic. Accurate sub-models for the locked differential, the braking system, and the centrifugal clutch have been included as well.
\begin{figure}[ht]
    \centering
        \includegraphics[clip, trim=0cm 0cm 0cm 0.5cm, width=1.0\columnwidth]{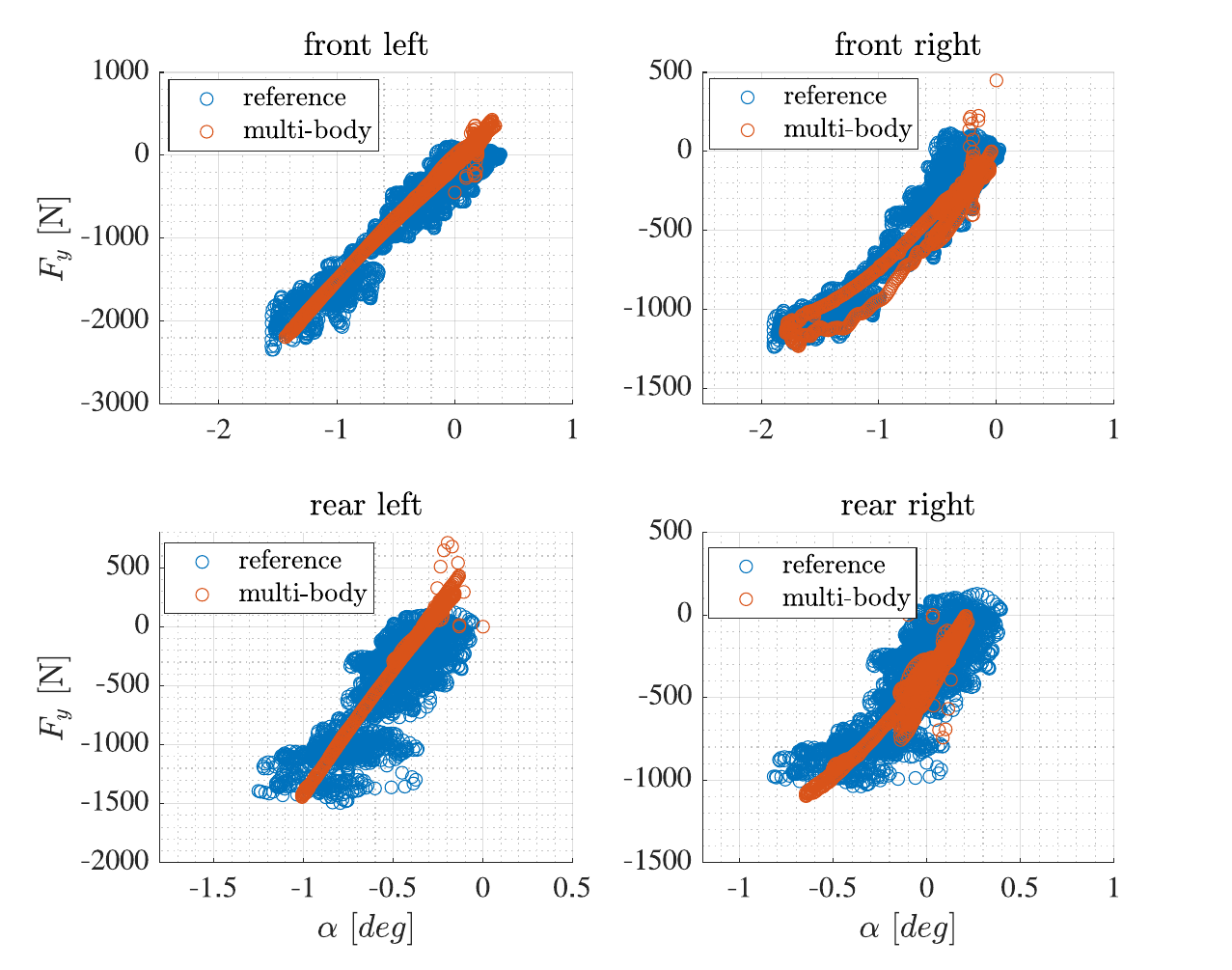}%
    \caption{A comparison between experimental data and the multibody model is presented. The slip angle is plotted on the x-axis, and the estimated wheel force is shown on the y-axis. It can be observed that the inner wheels (right side), where there is less vertical load acting, are working in the non-linear region of the tire.}
    \label{fig:dymola-tires}
\end{figure}
 \begin{figure}[ht]
     \centering
         \includegraphics[clip, trim=0cm 0cm 0cm 0.7cm, width=0.95\columnwidth]{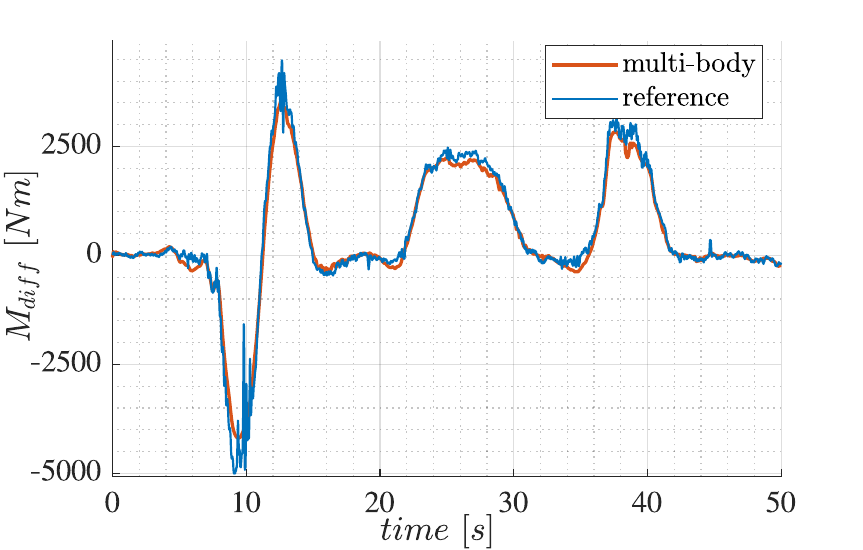}%
     \caption{Comparison between the experimental measurement of the yaw moment due to the locked differential (through slip ratio calculations), used as a reference (in blue), and the estimation from the multi-body model (in orange). The noisy reference is related to the accuracy of the estimation and the highly nonlinear nature of the measurement. The higher the curvature, the greater the resisting yaw moment to turn in. Here, the central section of the Monza track is shown, including Variante 2 and the Lesmo turns.}
     \label{fig:dymola-diff}
 \end{figure}

As can be seen in Figures \ref{fig:dymola-tires} and \ref{fig:dymola-diff}, the overall accuracy is good enough for the objective, thus the model is used as ground truth for calibrating the simplified model and for testing the control algorithms.

\subsection{Locked Differential Tricycle Model}
\label{tricycle_model}

For what concerns the front axle, the tricycle formulation is derived from the classic single-track model expressed in curvilinear coordinates, while the rear axle is derived from the double-track model. The final structure is shown in Figure \ref{fig:tricycle_model}. The new state-space model is described by the state vector 
$\Tilde{x} = [s; n; \mu; v_x; v_y; r; \delta; D]\,$ and the input vector $\Tilde{u} = [\Delta\delta; \Delta D]\,,$
where $s$, $n$, and $\mu$ represent the progress along the path, the orthogonal deviation from the path, and the local heading. The motion equations are expressed as the derivative of the longitudinal $v_x$ and the lateral $v_y$ velocities and the yaw rate $r$.
$\delta$ represents the steering angle and $D$ is the desired longitudinal acceleration.
$\Delta\delta$ and $\Delta D$ represent the derivatives of the inputs. The whole set of equations can be found in \cite{raji}.

\begin{figure}[ht]
    \centering
    \rotatebox{90}{%
        \includegraphics[clip, trim=8.7cm 1.0cm 7.5cm 1.5cm, width=0.64\columnwidth]{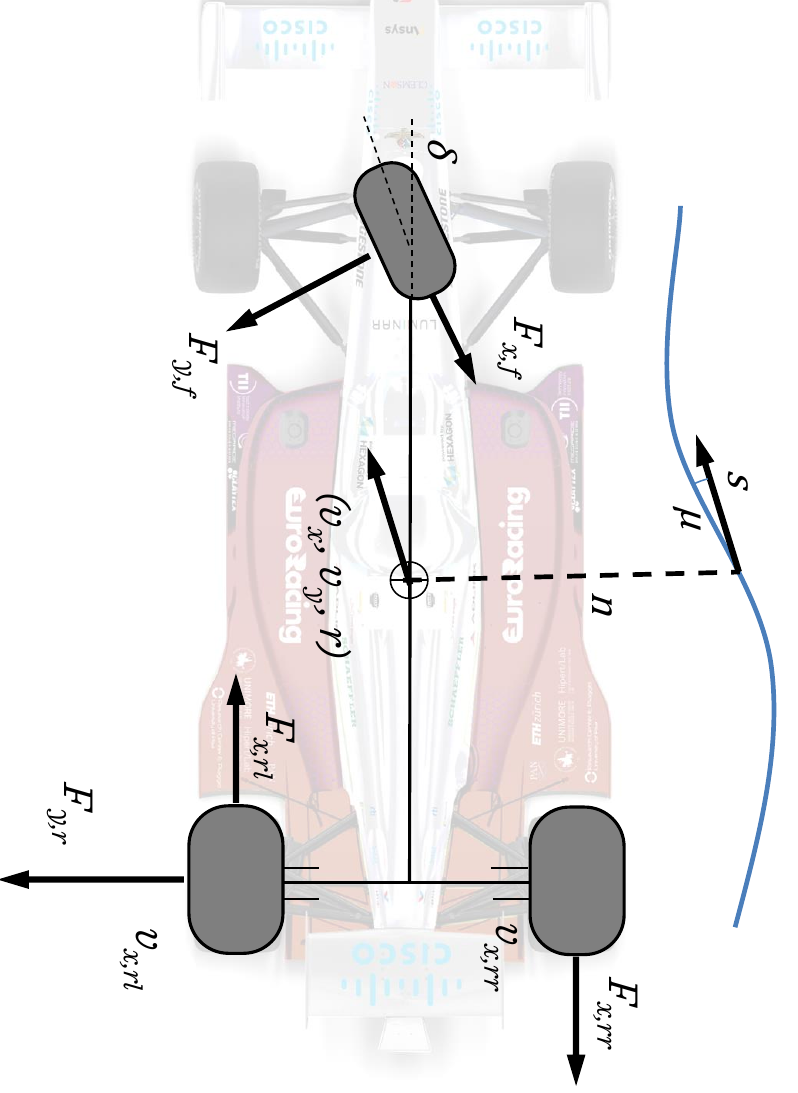}%
    }
    \caption{Scheme of the three-wheels model, in curvilinear coordinates, included in the MPC.}
    \label{fig:tricycle_model}
\end{figure}

In the motion equations, we included the longitudinal forces required for traction and braking $F^{cmd}_{xf}$ and $F^{cmd}_{xr}$, and the contribution of the longitudinal forces on the total yaw moment, $M_\textit{diff}$. The set of equations is presented here:

\begin{small}
\begin{subequations}
\begin{align}
    \label{eq:model:motion}
        \dot{v}_x =& \, \frac{\scalebox{0.8}{1}}{\scalebox{0.8}{\textit{m}}}\big(F^{cmd}_{xr} - F_{d}- F_{yf} \sin(\delta) + F^{cmd}_{xf} \cos(\delta) + m v_y r\big)\,,\\
        \dot{v}_y =& \, \frac{\scalebox{0.8}{1}}{\scalebox{0.8}{\textit{m}}}\big(F_{yr} + F_{yf} \cos(\delta) + F^{cmd}_{xf} \sin(\delta) - m v_x r\big)\,,\\
        \dot{r} =& \, \frac{\scalebox{0.8}{1}}{\scalebox{0.8}{\textit{I}}_z}\Big(M_\textit{diff} + l_f \big(F_{yf} \cos(\delta) + F^{cmd}_{xf} \sin(\delta)\big) - l_r F_{yr}\Big)\,,
\end{align}
\end{subequations}
\end{small}where $m$ and $I_z$ are the mass and the inertia of the vehicle, and $l_f$ and $l_r$ are the front and rear wheelbase. $F_{yf}$ and $F_{yr}$ are the lateral forces generated in the interface between the road and the tires, and $F_d$ is the aerodynamic drag. The rolling resistance is included in the longitudinal equilibrium as well. From now on, the index $i^{th}$ is used to identify the axle (front, rear), while the index $j^{th}$ identifies the vehicle side (left, right).

The yaw moment is derived from the rotational equilibrium of the rear axle in the Z-axis direction:
\begin{align}
        M_{\textit{diff}} =& \frac{1}{2}(F_{x,rr} - F_{x,rl}) \cdot t_r.
\end{align}
where $t_r$ represents the rear axle track. The longitudinal forces include both the contribution of the locked differential and the traction dynamics: $F_{x,rj} = F'_{x,rj} + F^{cmd}_{x,rj}$.
The longitudinal forces experienced during vehicle coasting are estimated using the non-linear Pacejka tire model:
\begin{align}
        F'_{x,rj} =& \, D_{x,rj} \, \sin\Big(C_{x,r} \tan^{-1}(B_{x,r} \, k_{x,rj})\Big)\,
\end{align}
$D_{x,rj}=\mu_r F_{z,rj}$, $B_{r}$ and $C_{r}$ are the macro-parameters of the magic formula \cite{pacejka} and $F_{z,rj}$ represents the vertical load on the $j^{th}$ wheel, as will be shown in \ref{Fzrj}. The slip ratio is expressed as
\begin{align}
        k_{x,rj} =& \, \frac{v_x - v_{x,rj}}{v_x}\,
\end{align}
and depends on the ideal speed of the rear wheels, taking into account the vehicle's geometry and the track's layout:
\begin{align}
        v_{x,rj} = r \cdot \left(R \pm \frac{t_r}{2}\right)\,. \label{vx,r}
\end{align}
In the equation \ref{vx,r}, $R$ is the turning radius.
To account for the traction forces, we have included a straightforward formulation based on the commanded acceleration. We defined $F^{cmd}_{x,r} = f(D)$, which is distributed between the rear wheels through the lateral load transfer. In this way, we can check for a possible nose-in during the phase of turn-exit.

Considering the tricycle structure, it proved necessary to include a model for estimating the lateral load transfer and, consequently, to distribute the vertical load on each wheel. Hence, the standard model based on the roll axis \cite{guiggiani} has been included. This has been employed exclusively for the rear axle, as the axle effective characteristic is used for the front one.
The vertical load on each wheel is computed:
\begin{align}
        F_{z,rj} =& \frac{1}{2} F_{z,r} \pm \Delta F^{lat}_{z,r} \, \label{Fzrj} 
\end{align}
where the global load on the rear axle $F_{z,r}$ is the sum of the static load $F^0_{z,r}$, the aerodynamic effect $F^{aero}_{z,r}$ \cite{raji} and the longitudinal load transfer $\Delta F^{long}_{z,r}$ contributions \cite{guiggiani}. 
The lateral load transfer is estimated with the relation:
\begin{align}
\label{eq:latLoadTransf}
        \Delta F^{lat}_{z,r} = F_{y,r} \frac{h_r}{t_r} + \frac{M_y}{t_r} \frac{K_r}{K_{tot}},
\end{align}
where $h_r$ is the height of the roll axis at the rear axle, $K_r$ and $K_{tot}$ are the rear axle roll stiffness and the total roll stiffness. $M_y = m \cdot a_y \cdot q$ and $F_{y,r}$ represent the moment and force generated by the inertia on the roll axis. To estimate the latter, a rotational equilibrium of the whole vehicle on the Z-axis is performed, from which it follows that:
\begin{align}
    \begin{split}
        & F_{y,r} = \, (m \cdot a_y \cdot l_f + M^0_\textit{diff}) \cdot \frac{1}{l} \label{Fyr}
    \end{split}
\end{align}
Here, $l$ and $q$ represent, respectively, the wheelbase and the Z-distance between the center of gravity and the roll axis. $M^0_\textit{diff}$ represents the yawing moment calculated by the model in the previous integration step and $a_{y}$ is calculated starting from the motion field.

The employment of this simplified roll-axis model for the load transfer estimation is an acceptable approximation since the roll motion of an open-wheel vehicle is negligible. This module allows the model to improve the accuracy in the calculation of the longitudinal forces, and consequently the prediction of the vehicle's pose.

The longitudinal forces are also used to account for the combined slip on the tires. The formulation is based on a friction ellipse, 
through which a weight factor $G_{y,ri}$ is computed as in \cite{raji}. Likewise, the weight factor is defined for the front axle.

\subsection{Model Stability Analysis}
\label{stability_analysis}
Representing a racecar with a mathematical model always requires special care. This is because the parameters of the model, especially those concerning the tires, can make the differential equation system very stiff and subject it to numerical instability if the integration method (or the step size) is not carefully chosen.
The original setup proposed for oval tracks uses the numerical method Runge-Kutta 4 (RK4) and a time-step $h = 0.04s$ \cite{raji}, which guarantees the stability only for longitudinal speed greater than $28$ $m/s$ on the single-track, and $32$ $m/s$ on the three-wheels model. This is not suitable for a road course track like Monza, where the speed can reach a minimum of $12$ $m/s$ (i.e. Variante 1).
To guarantee the numerical stability of the dynamic model at low speeds, it has been chosen an integration step size small enough to include the poles of the linearized system within the stability region defined by the numerical method \cite{Fish}.

The overall results of the stability analysis for the tricycle model are shown in Figure \ref{fig:tricycle_stability}.
An optimal value of time-step  $h = 0.008 s$ was chosen to cover all significant speed ranges expected on the track, i.e. ${v^{min}_x} = 8 m/s$. Below this threshold, the kinematic model written in the dynamic model states is used to avoid numeric oscillations on the solution. The two models are blended similarly to how it has been presented in \cite{amz}. 

In the first steps of the development and the experimental testing, the chosen integration method has been the explicit Euler, which provides better computation times compared to higher-order methods (i.e. RK4). As described in Section \ref{results_sec}, this choice has been changed in the final tests.

\begin{figure}[ht]
	\centering
	\includegraphics[clip, trim=0cm 0cm 0cm 0cm, width=0.7\columnwidth]{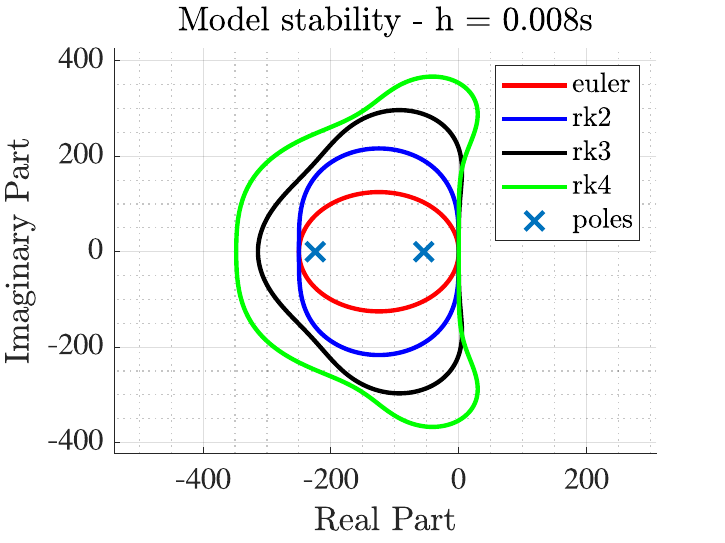}
    \caption{Poles for the three-wheels model at $v_x = 8 m/s$. The proposed step size for the integration allows the system poles to lay within the stability region of the explicit Euler method.}
	\label{fig:tricycle_stability}
\end{figure}

\section{MPC PROBLEM FORMULATION}
\label{mpc_sec}

The model in curvilinear coordinates described in Section \ref{tricycle_model} is used in an extension of the MPC presented in \cite{raji}. The main differences from the previous work are:
\begin{itemize}
\item A velocity tracking weight $q_v$ instead of a slack variable to follow more strictly the velocity reference.
\item A weight on the yaw rate $q_r$ which aims to mitigate potential oscillations caused by bumpy portions of the track or disturbances in the steering actuation.
\item Additional independent variables, and an improvement of the model discretization.
\item The inputs are the derivative of the steering angle and the requested longitudinal acceleration instead of using the throttle and brake commands. 
\end{itemize}
The cost function is formulated as:
\begin{small}
\begin{equation}
   J_\textit{MPC}(x_t, u_t) = q_n n_t^2 + q_\mu \mu_t^2 + q_v v_t^2 + q_r + u^T Ru + B(x_t)\,
\end{equation}    
\end{small}
where $q_n$ and $q_\mu$ are path following weights. The regularization term $B(x_k) = q_\textit{B} \alpha_r^2$ penalizes the rear slip angle, while $u^T Ru$ is a regularizer on the input rates where $R$ is a diagonal weight matrix.

The MPC problem is formulated as
\begin{small}
    \begin{subequations}
        \begin{align}
        \min_{X, U} &\ \ \sum_{t=0}^{T} J_\textit{MPC}(x_t, u_t) \\
        s.t. &\ \ x_0 = \hat{x} \,,\\
        &\ \ x_{t+1} = f_t^d(x_t, u_t)\,, \\
        &\ \ x_t \in X_{track} \quad x_t \in X_\textit{ellipse}\,,\\
        &\ \ a_t \in \bm A, \, u_t \in \bm U, t = 0,\dots,T.
        \end{align}
    \end{subequations}
\end{small}
where $X=[x_\textit{0}, ..., x_\textit{T}]$, and $U=[u_\textit{0}, ..., u_\textit{T}]$. $X_\textit{ellipse}$ represents a friction ellipse constraints, and $X_\textit{track}$ constrains the lateral deviation $n$ ensuring that the vehicle stays on the track.
$\hat{x}$ is the current curvilinear state and T is the prediction horizon.
$\bm{A}$ and $\bm{U}$ are box constraints for the physical inputs $a = [\delta; D]$ and their rate of change $u$.

\subsection{Locked Differential Model Progression}


In the definition of the optimization problem, the critical issue is advancing the prediction without the ability to track the evolution of $M_\textit{diff}$ with respect to the previous iteration. which is essential for estimating the correct value of the load transfer and the yaw moment itself. One solution could be the introduction of an additional variable to the state vector but this would increase the dimension of the optimization problem raising the computational time. 
Therefore, the $M^0_\textit{diff}$, which is the initial yaw moment for each step time $t$ in the control horizon, is calculated (with the same formulation as in Section \ref{vehicle_modeling}) before the execution of the control optimization, using the states from the latest MPC prediction horizon, and included as an independent variable. This is an acceptable simplification considering that having a stable model, a well-posed optimization problem, and a high-frequency running control scheme, the prediction of the MPC should not produce any abrupt change compared to the previous one.

\subsection{Micro-steps Model Discretization}

The MPC is executed in real-time at a frequency of 100Hz and a time horizon of 2.6s with a sampling time of $\Delta t$ = 40ms. In the previous version \cite{raji}, the model was discretized in time $f_t^d(x_t, u_t)$ with the same sampling time of the optimization:
\begin{equation} x_{k+1}=\Phi_{k}(x_{k},\ u_{k}, \Delta t), \end{equation}
where $\Phi_{k}$ is the chosen integrator (Euler or RK4).
However, this does not adhere to the outcome of the stability analysis presented in Section \ref{stability_analysis}. Therefore, a micro-step discretization approach has been applied
\begin{equation} \Phi_{k}(x_{k},\ u_{k}, \Delta t) = \underbrace{\Phi_k(\ldots \Phi_k}_{\text{$\Delta t / h$ times}}(x_k, u_k, h), u_k, h))\end{equation}
where $h$ indicates a chosen sampling time divisor of $\Delta t$, in our case $h$ = 8ms. This can also be seen as performing $\Phi_{k}$ with $\Delta t / h$ step sizes of $h$ and picking the resulting $x_{k+i}$ at each $\Delta t$.

\section{PLANNING}
\label{planning_sec}

\subsection{Global Trajectory}

To generate an optimal race line we used the global planner presented in \cite{raji}. Despite its effectiveness in producing an optimal path, the generation of different speed profiles is not trivial and rapid.
For this reason, a tool has been designed to compute and optimize the velocity profile for the generated path. Its primary objective is to find the most efficient velocity profile that minimizes lap time while respecting the specified constraints.

The first constraint pertains to the lateral acceleration,
\begin{equation}
    v^2_x \cdot \rho < a^{max}_{y}(v_x),
\end{equation}
where $v_x$ represents the vehicle speed along the path, $\rho$ is the curvature at a particular point, and $a^{max}_{y}$ is a function that defines the maximum allowable lateral acceleration for a given speed. The values for the maximum lateral acceleration are obtained from a ramp-steer maneuver simulated in the multi-body model presented in \ref{dymola}.

The second and third constraints focus on the negative and positive longitudinal acceleration. 
\begin{equation}
    a^{max}_{x+}(v_x) > \frac{\Delta v_x^2}{2 \Delta s} > a^{max}_{x-}(v_x) \label{axBounds}
\end{equation}
The constraints are determined by the rate of change of the velocity $\Delta v^2_x$ divided by twice the rate of change of distance $\Delta s$. The equation \ref{axBounds} states that this value must be greater than $a^{max}_{x-}$ and lower than $a^{max}_{x+}$, which are respectively the maximum permissible deceleration and acceleration for a given speed. The evaluation of the negative acceleration limit is performed by exploiting the braking diagram of the vehicle, accounting for brake balance set-up, longitudinal load transfer, and aerodynamic effects. While the longitudinal limit is found by considering both the limitation in the powertrain and the tire grip. 


\subsection{Longitudinal Planner}

The Longitudinal Planner dynamically computes speed, acceleration, and jerk profiles based on the real-time status of the vehicle. This module is based on a Linear Model Predictive Control (LMPC) framework, following the standard formulation: 
\begin{small}
\begin{subequations}
  \begin{align}
    \min_{X,U}  & J = \|x_\textit{T}-x_\textit{r,T}\|_P^2 + \sum_{t=0}^{T-1} \left( \|x_t - x_{r,t}\|_Q^2 + \|u_t\|_R^2 \right) \\
    \text{s.t.} \quad & x_{t+1} = Ax_t + Bu_t \\
    & x_0 = \hat x \\
    & u_t \in \mathcal{U}_t, \quad t = 0, 1, \ldots, T-1 \\
    & x_t \in \mathcal{X}_t, \quad t = 0, 1, \ldots, T
    \end{align}  
\end{subequations}    
\end{small}
where $X = [x_\textit{0}, \ldots, x_\textit{T}]$, $U = [u_\textit{0}, \ldots, u_\textit{T-1}]$ and $T$ is the prediction horizon length; $P$ and $Q$ are the terminal and stage cost weighting matrices on the deviation of the state $x$ from the reference $x_r$; $R$ is the control input cost weighting matrix; $\hat x$ is the current state; $\mathcal{X}_t$ and $\mathcal{U}_t$ are constraints on the state and input at time step $t$.
The prediction model is a double integrator including the longitudinal speed $v_x$ and acceleration $a_x$ in the state and the jerk $j_x$ as control input: 
\begin{align}
    x = & \begin{bmatrix}
        v_x & a_x
    \end{bmatrix}^{\top} , u = j_x\\
    A = & \begin{bmatrix}
        1 & T_s\\
        0 & 1
    \end{bmatrix} ,
    B = \begin{bmatrix}
        T_s^2/2 \\
        T_s
    \end{bmatrix}
\end{align}
where $T_s$ denotes the sample time.
Therefore, the goal is to plan feasible speed and acceleration profiles while minimizing the offline speed profile tracking error. Here, most of the design effort lies on the state and input constraints definition, i.e. $\mathcal{X}_t$ and $\mathcal{U}_t$ respectively.
Leveraging the preview of path curvature $\hat \rho_t$ and speed $\hat v_{x,t}$ over the upcoming prediction horizon, an estimate for lateral acceleration can be derived as $\hat a_{y, t} =\hat  \rho_t \hat v_{x,t}^2$.
This estimate is then used to compute the bounds over the prediction horizon on the longitudinal acceleration, according to the friction ellipse: 
\begin{small}
    \begin{align}
    a_{x,t}^{\star} = \bar a_{x}^{\star}(\hat v_{x,t})\sqrt{1 - \left(\dfrac{\hat a_{y,t}}{\bar a_{y}^{max}(\hat v_{x,t})} \right)^2}, \quad \star \in \{\text{\small $max, min$}\}
\end{align}

\end{small}
where $\bar a_{x}^{max}(v_x)$, $\bar a_{x}^{min}(v_x)$ and $\bar a_{y}^{max}(v_x)$ are the bounds on the longitudinal and lateral acceleration considering the vertical load only.
Eventually, given a desired maximum value of the lateral acceleration $a_y^{max} < \bar a_{y}^{max}(v_x)$, possibly time-varying $a_{y,t}^{max}$, we compute a bound on the longitudinal speed $v_{x,t}^{max}$, depending on the path curvature $\hat \rho_t$ as well.
The advantage of employing an online longitudinal planner lies in its capacity to guarantee that, whenever the vehicle significantly deviates from the predefined offline speed profile, the controller is provided with speed and acceleration profiles that are feasible concerning the vehicle's dynamics, thereby enhancing safety. Furthermore, this approach opens up the opportunity for online adjustment of the vehicle's longitudinal performance. 
    
\section{Low-level Longitudinal Control}
\label{lowlevel_sec}

The MPC, presented in Section \ref{mpc_sec}, produces longitudinal velocity $v^{mpc}_{x}$ and acceleration $a^{mpc}_{x}$, which are then converted into throttle and brake signals using a combination of two PI controllers and feed-forward controllers. 

To calculate the throttle feed-forward action, the current engine speed, $rpm$, and the torque target, $T_\textit{ref}$, are used to query a heuristically constructed look-up table. 
The torque is determined by the target longitudinal force $F^{ref}_{x}$, the radius of the rear wheel $r_w$, the gear ratio $\tau_i$ (with $i$ being the current gear), and the final drive $\tau_d$. The transmission efficiency $\eta_t$ is included as well:

\begin{equation} \label{torque_eq}
    T_\textit{ref} = \frac{F^{ref}_x \cdot r_w \cdot \tau_i \cdot \tau_d}{\eta_t}
\end{equation}

The force $F^{ref}_x$ indirectly incorporates the target velocity $v^{mpc}_{x}$ through its dependence on $F_{d}$ and $F_{roll}$. It also directly considers the acceleration target $a^{mpc}_x$ according to the relationship:
\begin{equation} \label{target_fx}
    F^{ref}_x = m \cdot a^{mpc}_x - F_d - F_{roll} + \frac{J_0 \cdot a^{mpc}_x} {r_w^2},
\end{equation}
where $J_0$ represents the rotational inertia of the mechanics and $F_{roll}$ is defined as in \cite{raji}.
To ensure that the target does not exceed the tire's longitudinal limit, the actual target used for throttle calculation is the minimum value between $F^{ref}_x$ and $F^{limit}_{x,r}$. The latter is determined using Pacejka's peak factor $D_{x,r}$, already mentioned in Section \ref{tricycle_model}.
However, when the force in \ref{target_fx} is negative, it is converted into a brake signal using the equation \ref{brake_ff}. Since braking dynamics are significantly simpler than powertrain dynamics, a linear equation can be employed with negligible error:
\begin{equation} \label{brake_ff}
    B = \frac{F^{tar}_x}{C_{b,f} + C_{b,f}} \cdot B^{max}
\end{equation}
$C_{b,f}$ and $C_{b,r}$ are the brake coefficients determined by the geometry and the characteristic of the braking system, e.g. the disk dimension, and parameterized on the maximum pressure that can be applied in the system, $B^{max}$. Likewise in the throttle calculation, the actual longitudinal force used in \ref{brake_ff} is the minimum between $F^{ref}_x$ and a new threshold $F^{limit}_{x,b}$. Unlike the traction force, braking dynamics involve both the front and rear axles, thus requiring the assessment of the maximum force on both axles. Indeed, the braking limit is proportional to the sum of both $D_{x,f}$ and $D_{x,r}$. 

Any errors in the feed-forward actions due to discrepancies in force estimation and other modeling errors were compensated for by the PI controllers, whose actions were added to the feed-forward ones.

\section{RESULTS}
\label{results_sec}
\subsection{Model Validation}
\label{model_validation}
\begin{figure}[b]
	\centering
	\includegraphics[clip, trim=0cm 0.6cm 0cm 0cm, width=1.0\columnwidth]{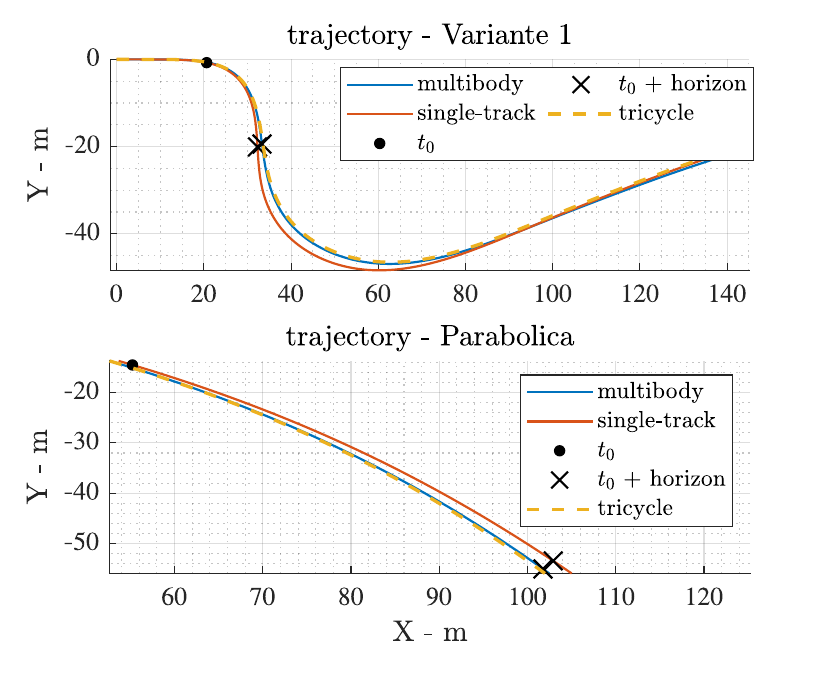}
    \caption{Monza circuit trajectory: the first one displays Variant 1, the tightest curve on the track, while the second one showcases Parabolica, one of the curves with the widest radius.}
	\label{fig:openloop}
\end{figure}
A comparison between the single-track model and the tricycle model in open-loop is presented in Figure \ref{fig:openloop}. The multi-body model serves as a reference. A time horizon of $2.6s$ (as the MPC horizon) is set to highlight position and yaw error over the entire time horizon.
It is evident that the prediction of the single-track model in tight-radius curves estimates a larger effective curvature radius, whereas in wide-radius curves the prediction would be the opposite. Concerning the single-track, this result is achieved through a proper tuning of the parameters of the constitutive equations, which represents a trade-off among all the curves of the Monza racetrack. As preliminary introduced in Section \ref{introduction}, improving the model to perfectly fit tight-radius curves will decrease performance in wide-radius turns, and vice versa.
The tricycle model enhanced by the contribution of the locked differential, using a single axle characteristic calibration, guarantees a better estimation over the entire track.

The results of the open loop scenarios are summarized in Table\;\ref{lateral-errors}, where $e_y$ and $e_{\psi}$ indicate the lateral ($m$) and heading (°) error respectively. The Lesmo turns are not included in the table as they did not produce significant data. 

\begin{figure}[b]
	\centering
	\includegraphics[clip, trim=0cm 0cm 0cm 0.8cm, width=1.02\columnwidth]{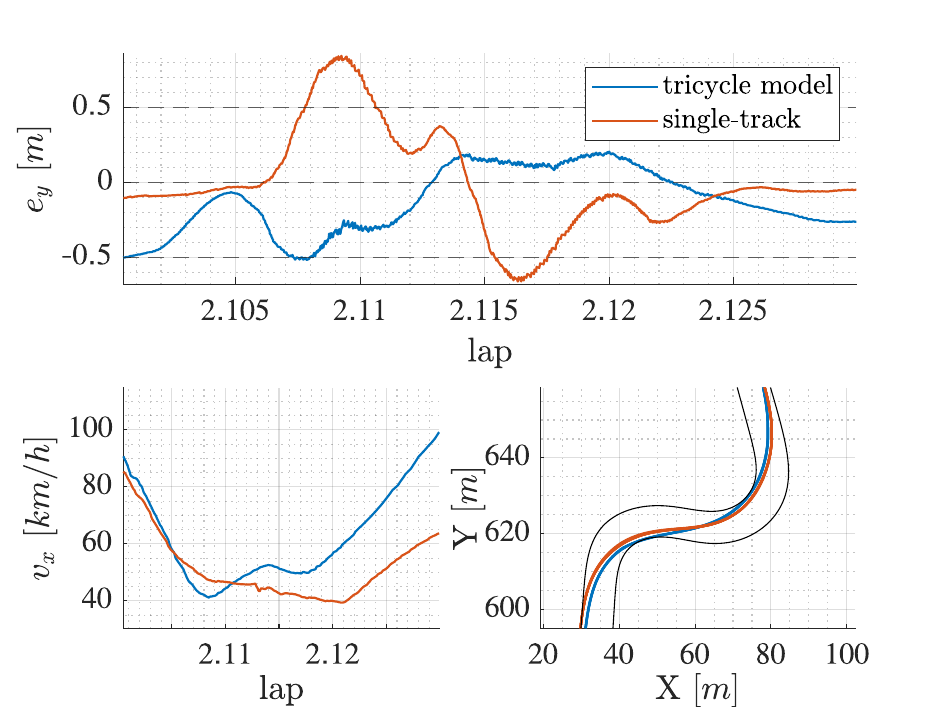}
    \caption{Comparison of the two models in the F1 circuit of Monza, Variante 1. See \url{https://youtu.be/jNc9D9T8inw?si=X4n-57CwFl1DXTEE} for the on-board video of the best lap.}
	\label{fig:Prelim-res-1}
\end{figure}
\begin{table}[h!]
\begin{center}
\caption{Lateral error $e_y$ and heading error $e_{\psi}$}
\small
\renewcommand{\arraystretch}{1.2}
\begin{tabular}{|c|c|c|c|c|c|c|c|c|}
\hline
& \multicolumn{2}{c|}{Variante 1} & \multicolumn{2}{c|}{Variante 2} & \multicolumn{2}{c|}{Ascari} & \multicolumn{2}{c|}{Parabolica} \\
\hline
& $st$ & $tric$ & $st$ & $tric$ & $st$ & $tric$ & $st$ & $tric$ \\
\hline
$e_y$ & 1.22 & 0.23 & 1.68 & 0.80 & 2.26 & 0.17 & 1.97 & 0.38 \\
\hline
$e_{\psi}$ & 4.14 & 0.86 & 4.98 & 2.19 & 3.98 & 0.18 & 1.87 & 0.41 \\
\hline
\end{tabular}
\label{lateral-errors}
\end{center}
\end{table}
\subsection{Preliminary Experimental Results}
\label{preliminary-res}

The on-track experimental results of the controller using the tricycle model, compared to the single-track model, are presented. Similarly to \cite{er-autopilot}, an Extended Kalman Filter (EKF) is used to filter all the signals of the vehicle and produce the states needed by the controller. 

The single-track model is defined by the kinematic model in the narrow-radius curves and by the dynamic model in the wide-radius curves as satisfactory and safe tuning could not be found for all the turns of the circuit under consideration. 
\begin{figure}[ht]
	\centering
	\includegraphics[clip, trim=0cm 0cm 0cm 0.7cm, width=0.95\columnwidth]{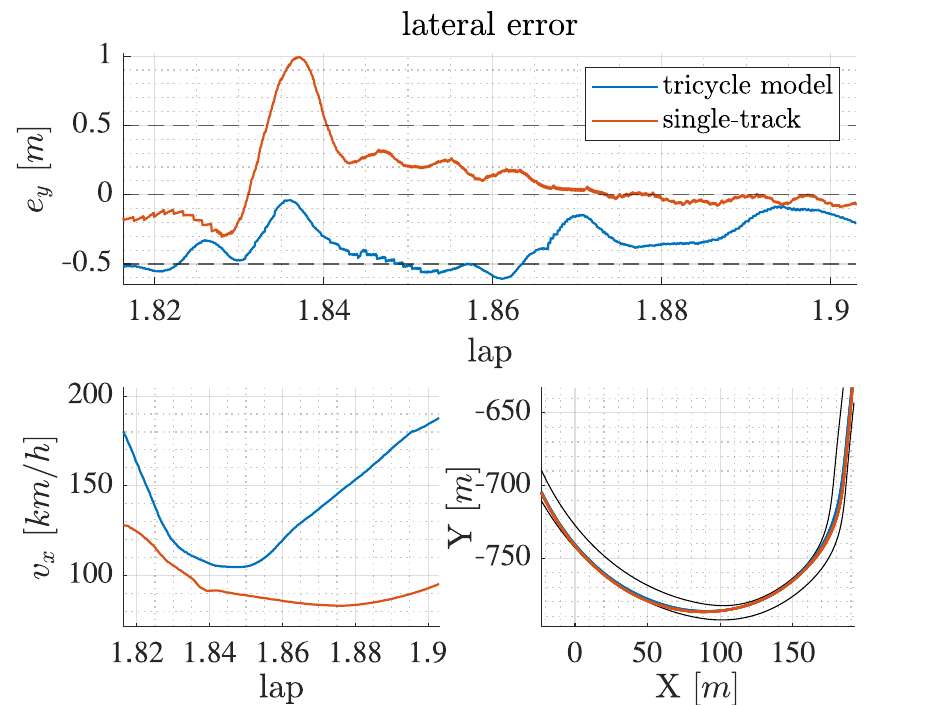}
    \caption{Comparison of the two models in the F1 circuit of Monza, Parabolica turn.}
	\label{fig:Prelim-res-4}
\end{figure}

Figure \ref{fig:Prelim-res-1} shows the lateral tracking error and longitudinal velocity with the two models. As expected, compared to the kinematic single-track model, the error using the tricycle model is significantly lower.
Similar results are presented in Figure \ref{fig:Prelim-res-4}, where the Parabolica turn is shown. With the dynamic single-track model, the error increased beyond $1$ $m$, forcing the car to slow down within safe limits. On the contrary, the tricycle model guaranteed a more consistent and stable behavior.
As can be noticed in Figure \ref{fig:Prelim-res-1} and Figure \ref{fig:Prelim-res-4}, both the models start the turn maneuvers with a not negligible lateral error. This has been caused mainly by the deterioration of the steering actuator performance at angles close to zero and the setup alignment of the front tires. Another limitation has been the usage of the Euler integrator. The decision was initially taken considering mainly the computational burden. However, this caused an inaccurate discretization when the tire utilization moved towards the non-linear region. 

\subsection{Final Results}
\begin{figure}[b]
	\centering
	\includegraphics[clip, trim=0.85cm 0.5cm 0cm 0cm, width=1.1\columnwidth]{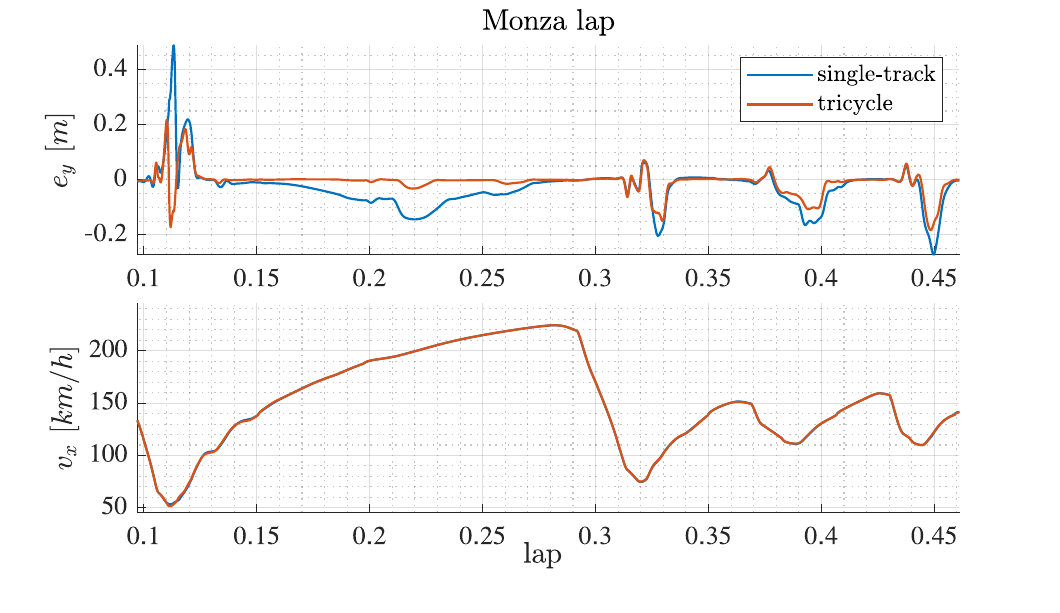}
    \caption{Comparison of the two models in a simulation on the 1st sector of Monza F1 virtual circuit, driven by the Model Predictive Control (MPC)}
	\label{fig:lap11}
\end{figure}
Despite the interesting insights of Section \ref{preliminary-res}, due to the described limitations and the different levels of accuracy in their tuning, we do not consider those results satisfactory enough for a correct comparison of the two models at their best. 
For this reason, and to demonstrate the performance of the proposed solution close to the very limit of the tire, we present the results in the Dymola multi-body model simulator. These have been produced using the RK4 integrator and after a further improvement of the models' fitting and minor changes on the cost tuning.

In Figures \ref{fig:lap11} and \ref{fig:lap12}, a comparison between the MPC using the two different models is shown.
\begin{figure}[b]
	\centering
	\includegraphics[clip, trim=0.85cm 0.5cm 0cm 0cm, width=1.1\columnwidth]{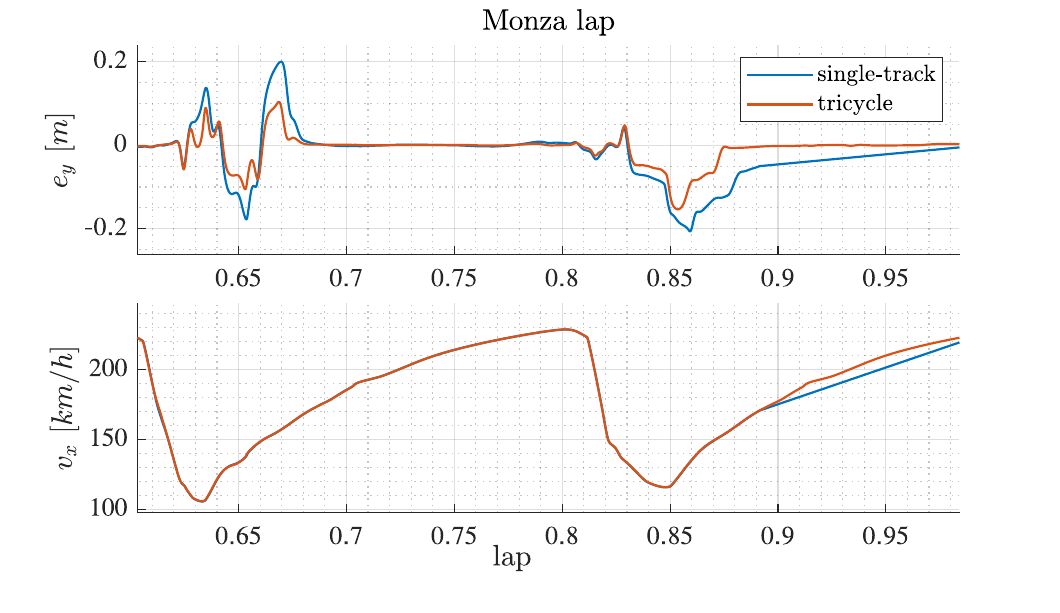}
    \caption{Comparison of the two models in a simulation on the 2nd sector of Monza F1 virtual circuit, driven by the Model Predictive Control (MPC)}
	\label{fig:lap12}
\end{figure}
Optimal tuning of effective axle characteristics has been defined for the single-track model to allow an acceptable error throughout the entire track. Furthermore, it should be noticed that differently from the results of the open-loop validation presented in Section \ref{model_validation}, due to the closed-loop feedback control the mismatch between the two models is reduced. The different behavior of the single-track model is still evident between slow and fast turns.
On the contrary, the tricycle model is capable of making sufficiently accurate predictions over the entire lap and ensuring good consistency in the controller behavior.

In Figure \ref{fig:fmu-cannone}, it is visible how the controller is able to exploit the non-linear region of the tire usage, demonstrating its effectiveness close to the vehicle handling limit. A colorbar has been included on the rear tires to highlight the effect of the longitudinal forces on the lateral characteristic. Here, positive combined slip values are associated with traction forces, whereas negative values correspond to braking forces.

\begin{figure}[ht]
	\centering
	\includegraphics[clip, trim=1cm 0cm 0cm 0cm, width=0.95\columnwidth]{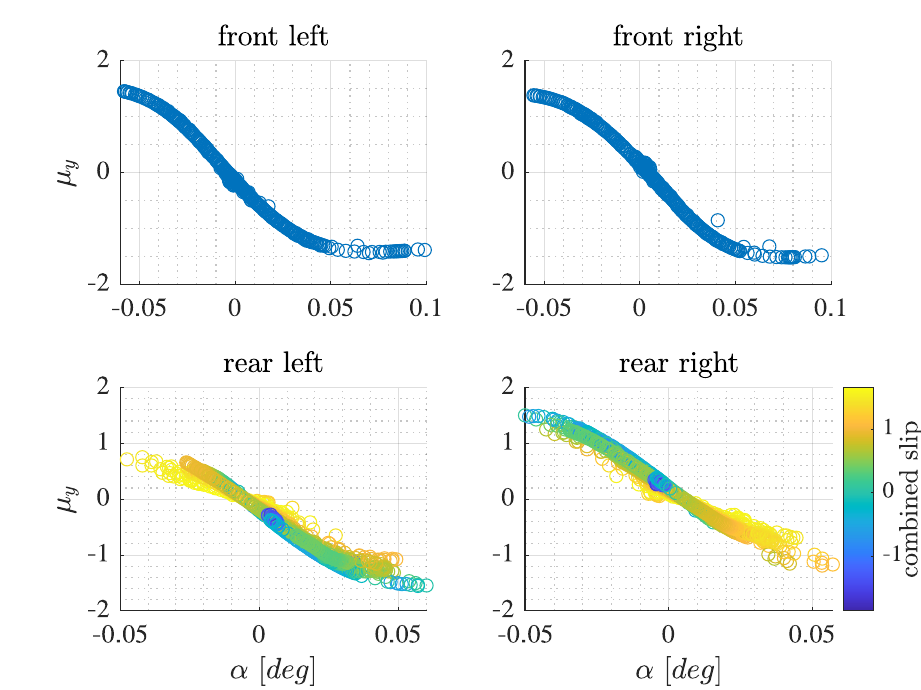}
    \caption{The characteristic curves of the tires exported from the telemetry of the multi-body model are displayed. It is interesting to note that the controller, with all its modules, can manage the exit phase of the curves at the traction limit.}
	\label{fig:fmu-cannone}
\end{figure}

\section{CONCLUSIONS}
\label{conclusions_sec}

A novel simplified vehicle model, defined as the tricycle or three-wheel model, and its integration into an MPC have been presented. Supported by an offline trajectory generation tool, a longitudinal planer, and a low-level longitudinal controller, the MPC with the proposed model demonstrated a reduction of the lateral error and a more consistent behavior compared to the classical single-track model in experimental tests on the Monza F1 racetrack. Further improvements in the model integration showed the capability of the controller to run the vehicle close to its limits on a high-fidelity simulator.

Lastly, it is evident that greater model accuracy can enhance the controller's prediction, improving both stability and performance. However, from the final results, it is clear that a simpler modeling approach, if well formulated, can still suffice for the purpose. This confirms that, depending on the objective and application, model selection has to be a trade-off. For an edge case application, the ideal raceline could be close to the edge of the track. In this scenario, it is important to reduce as much as possible the path tracking error to avoid touching the gravel or the track barriers which could lead to a crash or severe instability.

In future works, this model will be evaluated and extended to adapt to other vehicle configurations like the one with a limited-slip differential.

\addtolength{\textheight}{-12cm}   





\section*{ACKNOWLEDGMENT}

\label{ack_sec}
The authors would like to thank all the members of the TII Unimore Racing{\footnote{\href{https://www.tiiunimoreracing.com/}{https://www.tiiunimoreracing.com/}} team. We would also like to thank Claytex{\footnote{\href{https://www.claytex.com/}{https://www.claytex.com/}} for providing the VeSyMA Motorsports Library and for the technical support on the real-time execution of the Dymola model. Thanks to MegaRide{\footnote{\href{https://www.megaride.eu/}{https://www.megaride.eu/}} for their technical support.


\end{document}